# Formulation of the laws of conservation and non-conservation

**F Herrmann**

Abteilung für Didaktik der Physik, Karlsruhe Institute of Technology, D-76128 Karlsruhe Germany

E-mail: f.herrmann@kit.edu

**Abstract**

For each substance-like quantity, a theorem about its conservation or non-conservation can be formulated. For the electric charge e.g. it reads: Electric charge can neither be created nor destroyed. Such a statement is short and easy to understand. For some quantities, however, the proposition about conservation or non-conservation is usually formulated in an unnecessarily complicated way. Sometimes the formulation is not generally valid; in other cases only a consequence of the conservation or non-conservation is pronounced. A clear and unified formulation could improve the comprehensibility and simplify teaching.

## 1. Introdution

For each substance-like quantity one can formulate a proposition about its conservation or non-conservation [1-3]. Such statements are fundamental and they are at the same time easy to understand. However, when looking at the usual wording of such theorems, one often gets the impression that things are more complicated than they actually are. In some cases, the fact is formulated as if it were only valid under certain conditions; in others, consequences from conservation or non-conservation are described instead of making a statement about the quantity itself.

We will first, in section 2, remind what is meant by a substance-like quantity. In Section 3, the concept of conservation is distinguished from the concept of constancy. In section 4 we come to our actual topic: Common formulations of conservation or non-conservation are critically examined, one after the other, for energy, momentum, mass, electric charge, entropy, amount of substance and magnetic charge. In section 5 we discuss classes of processes in which quantities that are not generally conserved can be treated as conserved quantities. Finally, in section 6, consequences for teaching are discussed.

## 2. Substance-like quantities

The value of a physical quantity at a given instant of time refers to a geometric entity: a point, a line, an area or a region of space.

A temperature, a pressure, a mass density or an electric field strength refer to a point; voltage is defined for a line; force, electric current and power each refer to a surface area; mass, momentum, entropy and some others refer to a region of space. These latter quantities are called extensive quantities.

Extensive quantities are particularly important for the teaching of physics for two reasons:

1. They play a central role in the general structure of physics. Each of the classical subfields of physics has its own characteristic extensive quantity. For mechanics this is momentum, for electricity electric charge, for thermodynamics entropy. This results in a far-reaching analogy that can considerably simplify the teaching process [1-7].

2. The handling of these quantities is particularly simple. When dealing with an extensive quantity, one can make use of a powerful model: the substance model. One imagines the quantity as a measure of the amount of a substance or fluid, and one speaks about it as if one were speaking about this imaginary substance [8]. Because of this possibility the extensive quantities are also called substance-like quantities[1]. This way of treating a quantity is customary with the electric charge. One can recognize it by the formulations that are commonly used: the charge "flows", it can be "accumulated", "transferred", "concentrated", "distributed", "collected", there is "much" and "few" electric charge, etc.. This wording is a clear indicator that the substance model is being used.

This article deals with a special feature of the substance-like quantities: For each of them a statement about its conservation or non-conservation can be formulated. Thus, for example, momentum is conserved, entropy is not. For quantities that are not substance-like, there is no point in talking about conservation or non-conservation; temperature, pressure, force, magnetic field strength, for example, are neither conserved nor not conserved. The criterion "conserved or not conserved" simply does not apply in this case. (Note that being conserved does not mean that the value of the quantity is constant in time.)

Which are the substance-like quantities? Certainly they include mass, energy, momentum, electric charge and entropy; but also angular momentum (although it is not always possible to define an angular momentum density or a current density), magnetic charge (although it is hardly mentioned in today's textbooks), amount of substance (although most physicists consider it more as a "chemical" quantity), and finally quantities that in particle physics are called quantum numbers, such as the baryon number and the lepton number.

## 3. Conservation and constancy

Conservation and constancy are neighboring concepts; they are so close to each other that they are sometimes confused or used indistinctly.

In Hamiltonian mechanics, for example, it is often not distinguished between the terms "constant of motion" (also integral of motion) and "conserved quantity".

A constant of motion is a functional expression that is time independent and that is often rather unintuitive. However, not every quantity that is constant in time (under certain conditions) is a conserved quantity, and the value of a conserved quantity does not have to be constant in time for a given system. It is so only if an exchange of the quantity with the environment is excluded.

We therefore recommend, to make a clear distinction between "constant in time" and "conserved", whereby "being conserved" is the more important concept.

## 4. How to formulate conservation and non-conservation

Now to our actual subject: the formulation of conservation and non-conservation laws.

Actually, there should be no problem. If we use the substance model, i.e. if we apply the corresponding wording to a substance-like quantity $X$, then we can formulate the conservation of $X$ like this:

- $X$ cannot be created and cannot be destroyed.

If $X$ is not a conserved quantity, one of the following three statements would apply:

- $X$ can be created, but not destroyed.
- $X$ can be created and destroyed.
- $X$ can be destroyed but not created.

For the latter case, however, there is no example in nature.

In concrete terms, we would obtain the following propositions:

$$
\left.\begin{array}{l}
\text{Energy cannot be created and cannot be destroyed.} \\
\text{Momentum cannot be created and cannot be destroyed.} \\
\text{Angular momentum cannot be created and cannot be destroyed.} \\
\text{Electric charge cannot be created and cannot be destroyed.} \\
\text{Magnetic charge cannot be created and cannot be destroyed.} \\
\text{Entropy can be created, but not destroyed.} \\
\text{Amount of substance can be created and destroyed.}
\end{array}\right\} \quad (1)
$$

It may come as a surprise that this simple way of expressing conservation and non-conservation is rarely used in the textbook literature. Instead, there are often statements which are more difficult to understand and less universally applicable, and there are statements which do not even show that they refer to the conservation or non-conservation of a physical quantity.

In particular, there are two shortcomings. Instead of bluntly pronouncing the conservation:

- only consequences of the conservation are formulated;
- a procedure is described, which allows to check the conservation.

In the following, we shall critically discuss some popular formulations of the conservation or non-conservation for the most important quantities. Since we do not want to criticize any particular text book, we mostly do not give a source: We choose formulations in which the reader will probably recognize what he or she has heard or read in several occasions.

### 4.1 Energy

Often one refers to an energetically closed system:

- The total energy of a closed system remains constant.

The sentence makes a statement about a special case. In fact, the conservation of energy can be verified by looking at a closed system: One looks at the value of the energy for a while and realizes that it does not change. However, the conservation law also applies when the system is not closed. One can also check its validity with an open system: one simply has to take into account the flow of the incoming and the outgoing energy. One even does not need to consider any particular system at all to pronounce the theorem.

## 4.2 Momentum

The situation is similar regarding the law of momentum conservation. Here, too, one often refers to a closed system:
- In a closed system, the total momentum is constant.

First of all, it is an unnecessary detail to insist in the "total" momentum or the "sum of all momenta," as sometimes it is said. When speaking about the number of inhabitants of a city, one does not insist to mention the "total" number of inhabitants or the sum of all the subgroups of inhabitants either.

In addition, there is a more serious objection: Usually, before discussing momentum conservation, Newton's laws are treated. However, Newton's laws are nothing else than an expression of the conservation of momentum for special situations [1,4,5]. That means that either Newton's second law or the statement about momentum conservation, is not needed as a basic law.

## 4.3 Mass

Usually, it is taken for granted that mass is conserved in physical processes. Our way of speaking about mechanical processes does not encourage the student to ask wether mass is conserved or not. The suspicion that mass might not be conserved only appeared in the context of chemical reactions. But this was a question for the chemists only. And they answered it. Lomonossov discovered the theorem of mass conservation in chemical reactions.

So, for chemists mass conservation became an important theorem, for physicists it was a matter of course. However, it became interesting also for the physicists, when it was noticed in the context of nuclear reactions that the sum of the rest masses of the particles that participated in a reaction was not conserved. Thus for the physicists only the deviation from conservation was remarkable. Although this deviation is not a violation of the mass/energy conservation law if the energy-mass equivalence is taken into account, it has kept its disparaging name "mass defect" until today.

## 4.4 Electric charge

The treatment of the conservation of the electric charge is similar to that of mass. It is stressed that there are two types of charge, positive and negative. But usually, it is not judged important to mention that electric charge is conserved. It is assumed to be given from the outset.

## 4.5 Entropy

This is where the situation is most unpleasant. We begin with the modern and easily understandable formulation (1) of the second law:
- Entropy can be created, but not destroyed.     (a)

(Since in the following we shall refer to various different formulations of the second law, they will be labelled with a letter.)

It is generally admitted that entropy was defined or introduced by Clausius. However, already in 1908 Ostwald [9] had noticed that the old caloric of Carnot could be identified with Clausius' entropy. But this observation never affected the formulation of the second law.

The name "Second Law" was coined by Clausius, who formulated the theorem in several different ways, among them the following:
- Heat cannot spontaneously pass from a colder to a warmer body.     (b)

(Some of his other versions of the proposition are hard to understand today.)

Apparently one was not quite satisfied with this and the other formulations of Clausius. So more versions of the second law appeared. The one by Max Planck [10] has become particularly popular:
- It is impossible to construct a periodically working machine that does nothing else but lift a load and cool a heat reservoir.     (c)

Apparently, also this formulation was not considered satisfying, as is shown by the fact that in thermodynamic textbooks often more than one version of the second law is presented. In one German University textbook we found five different formulations listed in a table.

Let us revisit the two most popular versions of the second law, namely that of Clausius (b) and that of Planck (c).

What is the relationship between theorem (b) and the modern formulation (a)?

The "spontaneous" transition of heat from a warmer to a colder body is an irreversible or "dissipative" process, i.e. entropy is produced. If it were to run backwards, entropy would have to be destroyed. This is forbidden according to (a). Thus (b) follows from (a). What makes statement (b) somewhat awkward is that it describes only a particular consequence of the general theorem (a). Actually, theorem (a) has many other consequences, for example: Water only flows down the hill, never up; or electric charge only flows from high to low electric potential, never the other way round, momentum only goes from the body with the higher to that with the lower velocity; a chemical reaction only runs from high to low chemical potential. All of these processes are supposed to run spontaneously, i.e. irreversibly, just as the heat transport in Clausius's example.

Of course, from the fact that heat only flows from hot to cold one could deduce that water only flows downhill, but that would be rather clumsy. Why not use instead the general formulation (a), from which one can easily deduce all of the various special cases?

Also the formulation (c) of Planck can be viewed critically. We want to contrast it with another statement, which is also true:
- It is impossible to construct a periodically working machine that does nothing else than lift a load and discharge a reservoir of electric charge.

Notice that, because of the "nothing else", discharging means that the charge disappears, i.e. it is not transferred somewhere else. It is easily seen what causes such a machine to fail: electric charge cannot be destroyed. Accordingly, the machine mentioned by Planck is not possible because entropy cannot be destroyed. Planck's statement is true, but it says nothing about the possibility that entropy can be created. It is therefore not equivalent to proposition (a).

Finally, we found a formulation of the second law, which can only be described as an act of despair. Apparently also dissatisfied with the usual versions, the author of another German University textbook formulates the second law (in addition to three other formulation):

- There are irreversible processes.     (d)

An advantage of this statement is that it is easy to understand, and that it follows clearly from (a). However, (a) does not follow from (d), because statement (d) says nothing about who or what is the cause of the irreversibility.

In 1911 finally Jaumann [11] and his scholar Lohr [12] were able to formulate a local balance equation for the entropy. Such an equation is the clearest way to express our statement (a) mathematically. But it came too late. The inappropriate formulations were already deep-rooted in the textbook literature and therefore have survived.

### 4.6 Amount of substance

Amount of substance is a quantity that physicists know, but that they do not consider quite seriously. Since it only tells us the number of particles the feeling may be that it is not a physical quantity like the others. As a result, no statements are made about its non-conservation. It can be considered a collateral damage that the esteem for the associated intensive quantity, the chemical potential, is not very high [13].

### 4.7 Magnetic charge

Another oddity is the treatment of the magnetic charge. As far as teaching at school is concerned, it could be one of the most easy to teach conserved quantities. With the electric charge it has in common that it can assume positive and negative values. Its advantage over electric charge is that the effects one is dealing with are much stronger. The fact that charges can compensate and that charge is conserved, can be seen in an experiment much more clearly with the magnetic charge than with the electric charge.

Due to historical coincidences the quantity magnetic charge had a hard fate. It was gradually removed from the physics textbooks due to an epistemological misunderstanding. The argument is that a quantity magnetic charge does not exist because no magnetic monopole particles have been discovered. However, by eliminating the magnetic charge it became impossible, or at least very cumbersome, to describe the simple fact that the north and south poles of a magnet are equal and opposite. Equal in what? In the magnetic charge. Not even the fact that there are no magnetically charged particles can be expressed without the quantity, because one needs the magnetic charge in order to state that its value is zero for all the known particles.

Actually it is easy to formulate a local balance equation or continuity equation for the magnetic charge. The charge in this case is pure "bound charge" and the current is a pure displacement current.

## 5. Conservation under certain conditions

We have seen that certain physical quantities obey a conservation law: Energy, momentum, angular momentum, electric and magnetic charge. Others do not: entropy and amount of substance. (As far as mass is concerned, it simply merges into energy.)

The conserved quantities are particularly appealing to us, because their behavior is simpler than that of the non-conserved ones. If somewhere the value of such a quantity decreases, it must increase elsewhere.

Now one can have this advantage also with the non-conserved quantities, because there are situations, or more precisely: processes, in which a quantity that in principle is not conserved is neither produced nor destroyed, i.e. where it behaves like a conserved quantity. We want to investigate this for the two non-conserved quantities that were mentioned before.

### 5.1 Entropy

There are processes in which entropy production is very low. Low means that the production rate is small compared to the entropy currents that flow. (Both, production rate and current, are measured in the same unit.) In technical thermodynamics such processes are realized in good approximation.

In a thermal power plant, for example, the entropy currents at the inlet of the turbine and at the outlet are nearly equal. The entropy production is very low compared to the entropy current flowing through the turbine.

Thus, in this case entropy behaves like a conserved quantity. It is for this reason, that we are allowed to describe the steam turbine plant in the ingenious way in which Carnot described the steam engine by comparing it with a water wheel. The water current that flows towards the water wheel at high altitude is equal to that flowing away at low altitude. Similarly, the entropy current that flows into the steam engine (or a steam turbine circuit) at high temperature is equal to that flowing out at low temperature.

### 5.2 Amount of substance

In "reactions" the quantities of the participating substances change: in chemical reactions, in reactions with photons and phonons, in nuclear reactions and in reactions that are studied in particle physics. But there are traditionally established areas of science where the circumstances are such that the numbers of certain particles (the amounts of substance) do not change. They behave like conserved quantities. Chemistry is particularly important in this respect.

Since only excitation energies of some eV or some tens of eV play a role, chemical "elements" are not transformed into other elements. This means: As long as the methods of chemistry are employed, a conservation law applies to each chemical element. Thus, about 100 conservation laws are valid. Even if we do not say it explicitly, we always make use of it when establishing a reaction equation. There must be the same amounts (numbers of moles) of the chemical elements on both sides of the equation.

Apart from these two examples, entropy and amount of substance, there are also quantities of which we do not yet know whether they are conserved or not. An example is the baryon number. And we do not yet know what cosmology will tell us one day about energy conservation. In any case, there is no reason not to treat baryon number and energy as conserved quantities. After all, physics is not about finding ultimate truths, but about finding an appropriate description of the world.

## 6. The role of conservation laws in teaching

Laws of conservation are important laws of physics. They deserve to be treated in class. Their treatment is simple as soon as the substance character of a quantity has become clear. So the most important step in teaching is to make the students aware that certain physical quantities are "substance-like". The formulation of the principle of conservation or non-conservation is then only one more detail.

The above considerations have shown that for some physical quantities a statement about conservation or non-conservation occupies a prominent place in teaching (at school and university), while for others it is even not mentioned. The conservation of energy and momentum is discussed extensively; but no statement is formulated for the conservation of mass and electric charge. Does this mean that the conservation of mass and charge is not considered important? No. It means that their conservation is taken for granted from the outset.

This observation should make us think. For someone who does not know anything about physics, the conservation of mass and electric charge would be no less remarkable than the conservation of momentum and angular momentum or the producibility of entropy. Apparently it is only due to our teaching tradition, that we take certain facts for granted and others not. If we speak of the mass from the outset as if it were a representative of matter, and of matter as if it were something imperishable, it seems superfluous to pronounce a proposition of mass conservation. When we speak about the electric charge as we are used to do – the conductor sphere discharges, the charge flows off into the earth, etc.– we leave the learner no chance of doubting that the charge is conserved. However, the class room experiments by no means show it: Before, the charge was sitting on the conductor sphere, after it's gone. How should it be conserved? Only the language of the teacher has the effect that conservation is taken for granted. It is different, for example, with momentum: one gets to know it as a combination of $m$ and $v$. And it is not evident that this should be a conserved quantity. It is even not evident that it describes a characteristic of a body, that is independent of mass and velocity.

Here are our recommendations:

1. The same type of formulation should be used in all cases so that the common features of the quantities become clear.

2. One selects the formulations (1), i.e. one pronounces the conservation (or non-conservation) without hesitation. They are not described in terms of particular consequences.

3. Conditional conservation is also mentioned. Thus, a conservation law is formulated for the amount of substance in chemical reactions and for the number of baryons and the number of leptons in nuclear reactions.

And finally, something that we would recommend to someone who is disposed to move a little further away from the teaching tradition: the traditional formulation of Newton's laws is abandoned. Instead, momentum is introduced right at the beginning of mechanics as an independent, underived, conserved quantity [4].

## 7. Summary

For each substance-like quantity a theorem about its conservation or non-conservation can be formulated. Due to the historical development of physics, a tradition has developed to treat these important laws in very different ways. For some, conservation (or non-conservation) is hardly addressed, but simply taken for granted, for others it is formulated in an unnecessarily complicated way. We recommend an approach that is, firstly, uniform and, secondly, transparent. Any such statement about a quantity $X$ can be formulated as: "$X$ can (not) be created and (not) destroyed".

---

[1]The categories extensive and substance-like are not entirely congruent. Consider the quantity "monetary value". Here everyone uses the substance model; the quantity would therefore be substance-like. According to our criterion, however, the monetary value is not an extensive quantity, because it cannot be assigned to a region of space; there is also no density and no current density for the monetary value.